\documentclass[reprint, superscriptaddress, aps, prl]{revtex4-2}

\usepackage{times,xspace}
\usepackage{graphicx,graphics,color,epsfig}
\usepackage{amsmath}
\usepackage{amssymb}
\usepackage{amsfonts}
\usepackage[version=4,arrows=pgf-filled]{mhchem}
\usepackage{epstopdf}
\usepackage[T1]{fontenc}
\usepackage{newtxtext,newtxmath}
\usepackage{xcolor}
\usepackage{siunitx}
\usepackage{braket}
\usepackage{xcolor}
\usepackage{bbm}
\usepackage{color}
\usepackage{float}
\usepackage[colorlinks=true]{hyperref} 
\usepackage{tikz}
\definecolor{myred}{rgb}{0,0,0}
\definecolor{myhighlight}{rgb}{1,1,1}

\definecolor{myred2}{rgb}{0,0,0}
\definecolor{myhighlight2}{rgb}{1,1,1}

\definecolor{mplblue}{RGB}{136, 204, 238}  
\definecolor{mplred}{RGB}{204, 102, 119}  

\hypersetup{
    unicode=false,              
    pdftoolbar=true,            
    pdfmenubar=true,            
    pdffitwindow=false,         
    pdfstartview={FitH},        
    pdftitle={Tuning chiral anomaly in a Dirac semimetal via fast-ion implantation},                
    pdfauthor={Manasi Mandal},  
    pdfsubject={},              
    pdfcreator={},              
    pdfproducer={},             
    pdfkeywords={} {} {},       
    pdfnewwindow=true,          
    colorlinks=true,            
    linkcolor=blue,             
    citecolor=blue,             
    filecolor=magenta,          
    urlcolor=blue               
} 
\usepackage[normalem]{ulem}

\newcommand{\figref}[1]{Fig.~\ref{#1}}

\def\be{\begin{equation}}
\def\ee{\end{equation}}
\def\bea{\begin{eqnarray}}
\def\eea{\end{eqnarray}}

\makeatletter
\renewcommand*{\@fnsymbol}[1]{\ensuremath{\ifcase#1\or \dagger\or *\or \ddagger\or
   \mathsection\or \mathparagraph\or \|\or **\or \dagger\dagger
   \or \ddagger\ddagger \else\@ctrerr\fi}}
\makeatother

\begin{document}
\title{Tuning chiral anomaly signature in a  Dirac semimetal via fast-ion implantation}

\author{Manasi Mandal}
\thanks{These authors contribute to this work equally.}
\affiliation{Quantum Measurement Group, MIT, Cambridge, MA 02139, USA}
\affiliation{Department of Nuclear Science and Engineering, MIT, Cambridge, MA 02139, USA}

\author{Eunbi Rha}
\thanks{These authors contribute to this work equally.}
\affiliation{Quantum Measurement Group, MIT, Cambridge, MA 02139, USA}
\affiliation{Department of Nuclear Science and Engineering, MIT, Cambridge, MA 02139, USA}

\author{Abhijatmedhi Chotrattanapituk}
\thanks{These authors contribute to this work equally.}
\affiliation{Quantum Measurement Group, MIT, Cambridge, MA 02139, USA}
\affiliation{Department of Electrical Engineering and Computer Science, MIT, Cambridge, MA 02139, USA}

\author{Denisse C\'{o}rdova Carrizales}
\thanks{These authors contribute to this work equally.}
\affiliation{Quantum Measurement Group, MIT, Cambridge, MA 02139, USA}
\affiliation{Department of Nuclear Science and Engineering, MIT, Cambridge, MA 02139, USA}

\author{Alexander Lygo}
\affiliation{Materials Department, University of California, Santa Barbara, CA 93106, USA}

\author{Kevin B. Woller}
\affiliation{Department of Nuclear Science and Engineering, MIT, Cambridge, MA 02139, USA}
\affiliation{Plasma Science and Fusion Center, Massachusetts Institute of Technology, Cambridge, MA 02139, USA}

\author{Mouyang Cheng}
\affiliation{Quantum Measurement Group, MIT, Cambridge, MA 02139, USA}
\affiliation{Center for Computational Science and Engineering, MIT, Cambridge, MA 02139, USA}
\affiliation{Department of Materials Science and Engineering, MIT, Cambridge, MA 02139, USA}

\author{Ryotaro Okabe}
\affiliation{Quantum Measurement Group, MIT, Cambridge, MA 02139, USA}
\affiliation{Department of Chemistry, MIT, Cambridge, MA 02139, USA}

\author{Guomin Zhu}
\affiliation{Materials Department, University of California, Santa Barbara, CA 93106, USA}

\author{Kiran Mak}
\affiliation{Quantum Measurement Group, MIT, Cambridge, MA 02139, USA}
\affiliation{Department of Materials Science and Engineering, MIT, Cambridge, MA 02139, USA}
\affiliation{Department of Physics, MIT, Cambridge, MA 02139, USA}

\author{Chu-Liang Fu}
\affiliation{Quantum Measurement Group, MIT, Cambridge, MA 02139, USA}
\affiliation{Department of Nuclear Science and Engineering, MIT, Cambridge, MA 02139, USA}

\author{Chuhang Liu}
\affiliation{Condensed Matter Physics and Materials Science Department, Brookhaven National Laboratory, Upton, NY 11973, USA}

\author{Lijun Wu}
\affiliation{Condensed Matter Physics and Materials Science Department, Brookhaven National Laboratory, Upton, NY 11973, USA}

\author{Yimei Zhu}
\email[]{zhu@bnl.gov}
\affiliation{Condensed Matter Physics and Materials Science Department, Brookhaven National Laboratory, Upton, NY 11973, USA}

\author{Susanne Stemmer}
\email[]{stemmer@mrl.ucsb.edu}
\affiliation{Materials Department, University of California, Santa Barbara, CA 93106, USA}

\renewcommand{\thefootnote}{\fnsymbol{1}}
    
\author{Mingda Li}
\email[]{manasim@mit.edu}
\email[]{mingda@mit.edu}
\affiliation{Quantum Measurement Group, MIT, Cambridge, MA 02139, USA}
\affiliation{Department of Nuclear Science and Engineering, MIT, Cambridge, MA 02139, USA}
\affiliation{Center for Computational Science and Engineering, MIT, Cambridge, MA 02139, USA}

\begin{abstract}
Cd$_3$As$_2$ is a prototypical Dirac semimetal that hosts a chiral anomaly and thereby functions as a platform to test high-energy physics hypotheses and to realize energy efficient applications. Here we use a combination of accelerator-based fast ion implantation and theory-driven planning to enhance the negative longitudinal magnetoresistance (NLMR)—a signature of a chiral anomaly—in Nb-doped Cd$_3$As$_2$ thin films. High-energy ion implantation is commonly used to investigate semiconductors and nuclear materials but is rarely employed to study quantum materials. We use electrical transport and transmission electron microscopy to characterize the NLMR and the crystallinity of Nb-doped Cd$_3$As$_2$ thin films. We find surface-doped Nb-Cd$_3$As$_2$ thin films display a maximum NLMR around $B = 7$ T and bulk-doped Nb-Cd$_3$As$_2$ thin films display a maximum NLMR over $B = 9$ T--all while maintaining crystallinity. This is more than a 100\% relative enhancement of the maximum NLMR compared to pristine Cd$_3$As$_2$ thin films ($B = 4$ T). Our work demonstrates the potential of high-energy ion implantation as a practical route to realize chiralitronic functionalities in topological semimetals.
\end{abstract}

\maketitle

The Adler-Bell-Jackiw (ABJ) chiral anomaly, originally formulated in quantum field theory, describes the quantum violation of chiral charge conservation. In the context of Weyl semimetals, this anomaly manifests as a chiral charge imbalance: an effective pumping of charge carriers between Weyl nodes of opposite chirality in momentum space when collinear electric and magnetic fields are applied. This quantum effect breaks the classical conservation law for chiral current and leads to observable transport phenomena such as negative longitudinal magnetoresistance (NLMR). The realization of the chiral anomaly in condensed matter systems not only deepens our analogous understanding of high-energy particle physics but also opens pathways for functional applications that exploit its topologically protected and universal transport characteristics \cite{hirschberger2016chiral, li2016negative, pikulin2016chiral, zhang2016signatures, huang2015observation, parameswaran2014probing}.
If we could control the chiral anomaly, we could systematically study anomalous and non-local transport phenomena with potential applications to sensing, switching, and chiraltronic devices \cite{takiguchi2020quantum,park2022real,kharzeev2014chiral,ong2021experimental}. However, tuning the chiral anomaly remains challenging given the difficulty in controlling the Weyl node horizontal separation, vertical energy offset, and the entanglement of the nodes with trivial band contributions. 

Cadmium arsenide (Cd$_3$As$_2$) is an archetypal topological Dirac semimetal which becomes a Weyl semimetal when a magnetic field breaks time-reversal symmetry and lifts the four-fold degenerate Dirac nodes into two-fold degenerate Weyl nodes. Cd$_3$As$_2$ has a high electronic mobility of $10^6$ cm$^2$/(V s) at low temperatures \cite{liang2015ultrahigh}, a strong linear magnetoresistance \cite{liang2015ultrahigh}, an anomalous Nernst effect \cite{liang2017anomalous}, and an NLMR--a signature of a chiral anomaly--in its various physical forms, including nanostructures \cite{li2015giant, jia2016thermoelectric, li2016negative}. Cd$_3$As$_2$ thin films grown via molecular beam epitaxy (MBE) exhibit two-dimensional surface states and develop a band gap due to quantum confinement \cite{schumann2018observation}.

Here we measure an NLMR that persists for a higher range of magnetic field in Cd$_3$As$_2$ thin films after doping with niobium (Nb). We implant Nb ions using fast ion implantation \cite{mandal2024precise} into Cd$_3$As$_2$ thin film heterostructures grown via MBE to lower the Fermi level to the Dirac node. To observe an NLMR, the Fermi level should be located near the Dirac point. Fast ion-implantation is commonly used in semiconductors \cite{ziegler1985high, pearton1990ion} and nuclear materials \cite{yan2015effects, das2019recent} research but rarely applied to quantum materials.  We find that surface-doped Nb-Cd$_3$As$_2$ thin films display a maximum NLMR around $B = 7$ T and bulk-doped Nb-Cd$_3$As$_2$ thin films display a maximum NLMR over $B = 9$ T while maintaining crystallinity. The maximum NLMR is enhanced by over 100\% compared to pristine thin films ($B = 4$ T). This is in contrast to the typical observation of the suppression of quantum effects as disorder increases \cite{Vojta2019Disorder}.  This work shows that high-energy ion implantation can be used to implant relevant species across the surface and bulk of topological thin films and enhance macroscopic quantum effects without destroying crystallinity.

\begin{figure*}[!htbp]
    \centering
    \includegraphics[width=\linewidth]{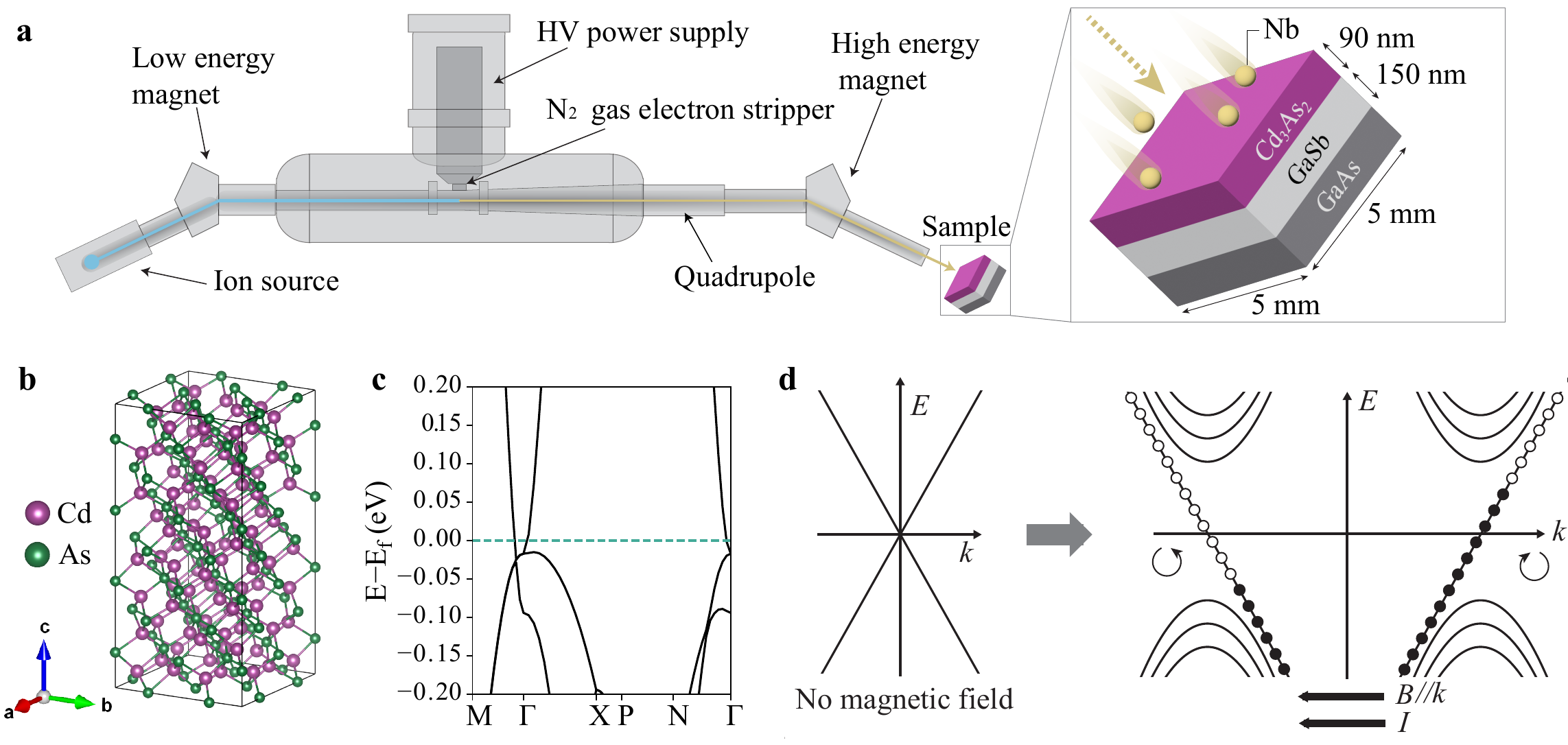}
    \caption{Fast-ion irradiation experimental setup, structural details, and Cd$_3$As$_2$ chiral anomaly illustration. \textbf{a.} Tandem ion accelerator schematic with Cd$_3$As$_2$ sample. An ion source produces an Nb ion beam, which is injected into the terminal using a low-energy magnet and focused by a magnetic quadrupole lens before being directed into the sample ($5$ mm $\times$ $5$ mm). The right side of (a) shows a magnified diagram of Cd$_3$As$_2$ (90 nm) deposited on a GaSb buffer layer (150 nm) on top of a GaAs substrate. The irradiating ion beam is perpendicular to the Cd$_3$As$_2$ surface. \textbf{b.} Crystal structure of Cd$_3$As$_2$ with 160 atoms per tetragonal unit cell. \textbf{c.} Dirac band structure of Cd$_3$As$_2$ near the Fermi level, which is indicated by a dashed line. The Fermi level lies above the Dirac node. \textbf{d.} Schematics of the band structure in a chiral anomaly. An intense $B$ field along a certain momentum separates the Dirac node as a result of broken time-reversal symmetry. The Weyl states are quantized into Landau levels (LLs) in the quantum limit. The $N$ = 0 LL has a gapless linear dispersion with slopes determined by $\chi$ denoted by clockwise and counterclockwise arrows. The magnetic field causes charge pumping between Weyl nodes, leading to a chiral charge imbalance as represented by the filled circles. This chiral current results in a negative longitudinal magnetoresistance. Adapted from Ref. \cite{xiong2015evidence}.
    }
    \label{intro}
\end{figure*}

\begin{figure*}[!htbp]
    \centering
    \includegraphics[width=\linewidth]{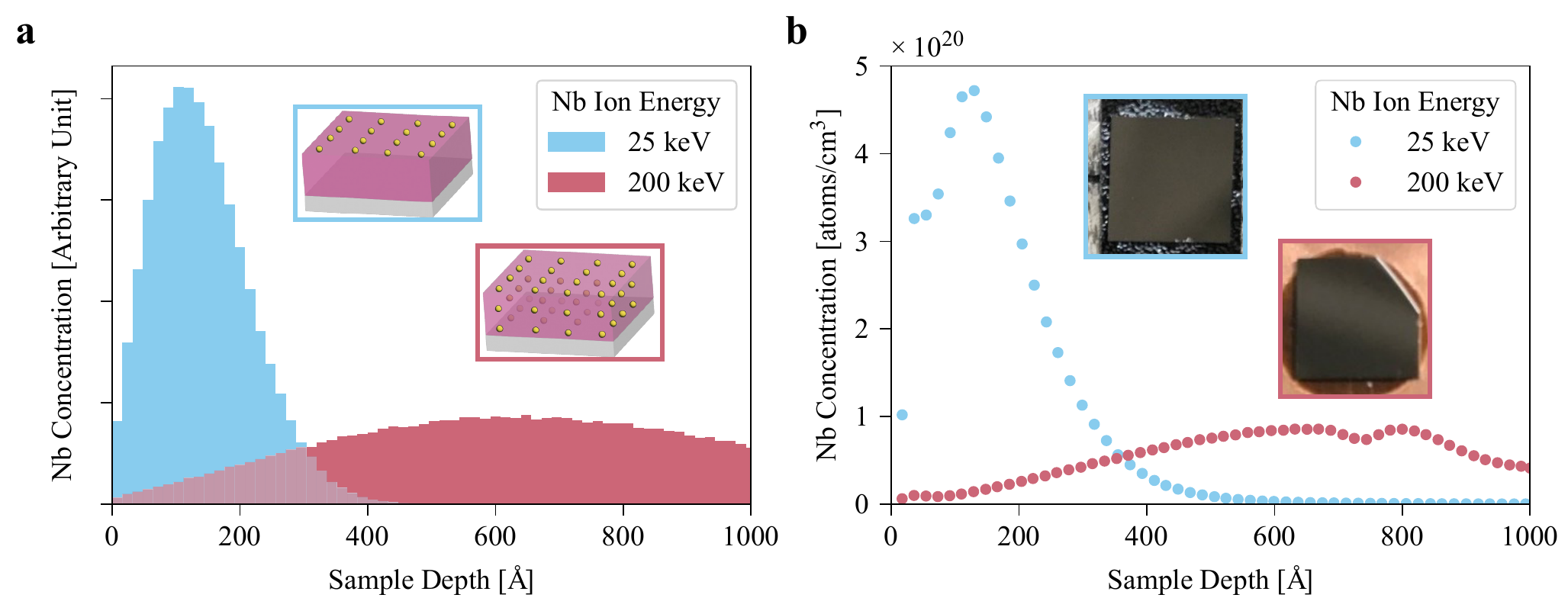}
    \caption{{Simulated and experimental profiles of Nb ion implantation in Cd$_3$As$_2$. \textbf{a.} SRIM/TRIM software simulated Nb ion implantation depth distributions for 25 keV, surface doped (blue), and 200 keV, bulk doped (pink), in Cd$_3$As$_2$ with corresponding illustrations of dopant concentration profiles as insets. The 25 keV surface-doped sample has the majority of implanted $^{93}$Nb$^{-}$ ions within 400 Å of the surface while the 200 keV beam results in bulk doping of $^{93}$Nb$^{+}$. \textbf{b.} Experimental Nb dopant concentration analyzed by Secondary Ion Mass Spectrometry Ref.\cite{SIMS} shows agreement with the calculated depth profiles. The samples were analyzed using an O beam for the best detection limits of Nb. The dip around $750$ Å in the bulk doped sample is potentially due to the emergence of an amorphous boundary layer between the {Cd$_3$As$_2$} and GaSb as shown in Fig.~\ref{fig3}. Insets are photographs of the samples.}}
    \label{fig2}
\end{figure*}

\begin{figure*}[!htbp]
    \centering
    \includegraphics[width=.8\linewidth]{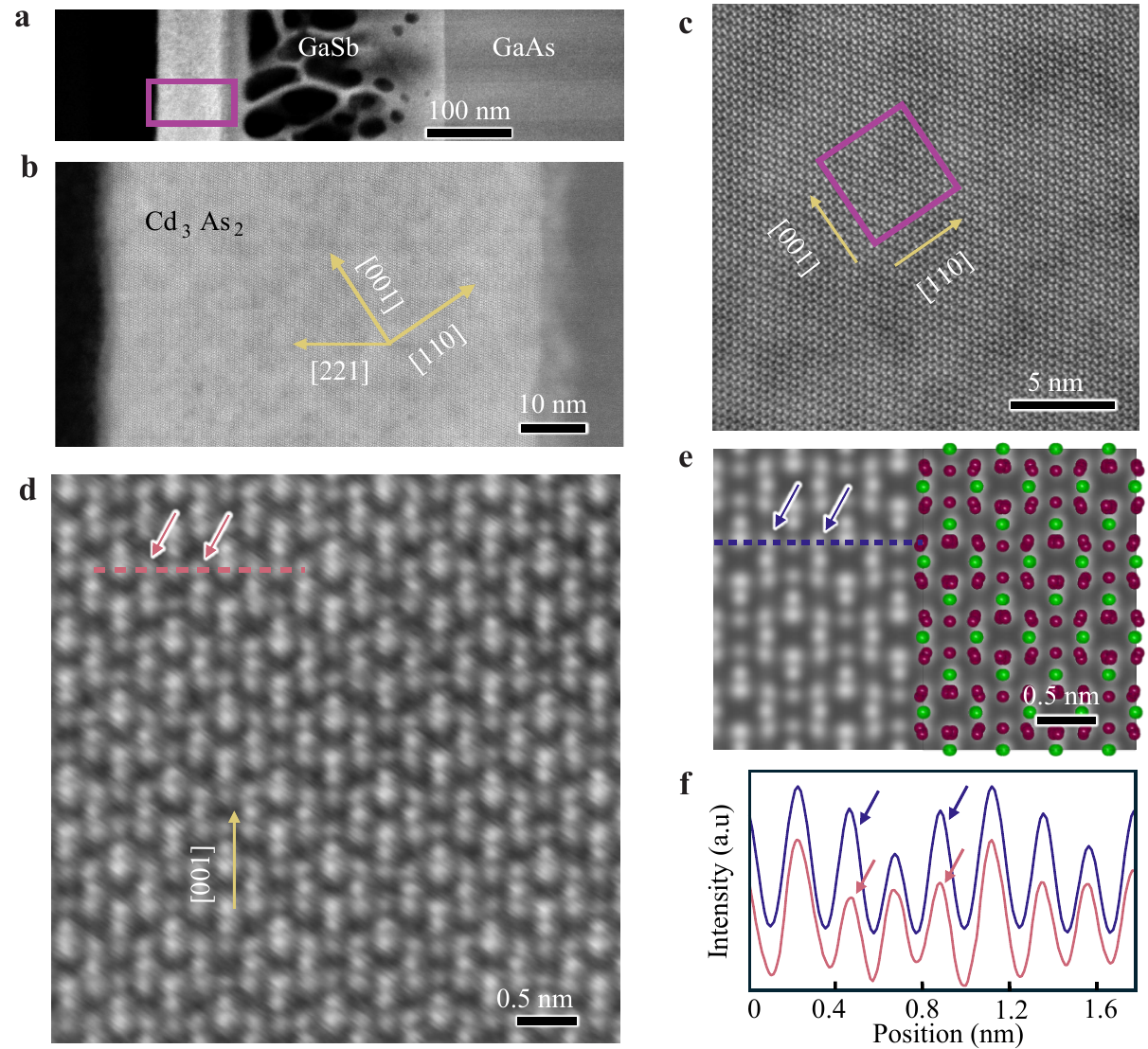}
    \caption{Scanning transmission electron microscopy (STEM) characterization of the Nb implanted Cd$_3$As$_2$ thin film with bulk doping. \textbf{a.} \textbf{b.} Low (a) and middle (b) resolution STEM images viewed along the [$1\bar{1}0$] direction of the Nb-Cd$_3$As$_2$ acquired by high-angle annular dark-field (HAADF). (b) is taken from the Nb-Cd$_3$As$_2$ thin film marked by the purple rectangle in (a). The heterostructure consists of a crystalline Cd$_3$As$_2$ film about 70 nm thick and an amorphous layer about 20 nm thick between the crystalline Cd$_3$As$_2$ and GaSb. Here, most of the GaSb layer is damaged or amorphized while the Cd$_3$As$_2$ layer remains structurally intact. \textbf{c.} High resolution STEM-HAADF image from the Cd$_3$As$_2$ film viewed along the [$1\bar{1}0$] direction. \textbf{d.} Digitally zoomed-in area marked by the purple square shown in (c). \textbf{e.} Calculated STEM-HAADF image based on perfect Cd$_3$As$_2$ crystal structure. The [$1\bar{1}0$] projection of the Cd$_3$As$_2$ structure is shown in the right with dark purple and green spheres representing Cd and As, respectively. f. Image intensity profiles from the scan lines in (d: pink line) and (e: blue line), respectively. Overall, the simulated image (e) is consistent with the experimental image (d), indicating that the intrinsic crystal structure of pristine Cd$_3$As$_2$\cite{zhu2023interface} is well preserved after the ion bombardment. A close inspection, however, shows that the relative image intensity in some atomic columns, e.g., those indicated by the pink arrows, deviates from the simulated image intensity as indicated by the blue arrows in (d,f). These relative disagreements in intensity indicate vacancies, likely induced by ion bombardment, in the Cd$_3$As$_2$ film.}
    \label{fig3}
\end{figure*}

To shift the Fermi level in Cd$_3$As$_2$ closer to the Dirac nodes, we used the General Ionex 1.7 MV Tandem Ion Accelerator at the Plasma Science and Fusion Center to implant Nb ions into the thin films. The schematic of the experiment is shown in \figref{intro}a. First, we loaded a niobium and graphite powder mixture (1:1) onto an ion source tip \cite{middleton1989negative}. Then, a cesium source bombards the powder mixture to sputter NbC$^-$, NbC$_2^-$, Nb, NbO$^-$ ions. For niobium, $^{93}$NbC$^{-}$ is the majority species produced. The niobium carbides and oxides are directed towards a low-energy magnet to select for Nb ions. The selected $^{93}$Nb$^{-}$ ions are extracted from the ion source at an energy of 25 keV. These ions are passed through the accelerator without charge exchange directly to the sample for implantation. For the 200 keV energy ions, N$_2$ gas is leaked into the high voltage terminal, stripping electrons from the Nb. The terminal potential was set to 87.5 kV, and the high energy magnet was tuned to steer 200 keV $^{93}$Nb$^{+}$ ions to the sample chamber. Nb acts as a dopant in Cd$_3$As$_2$ whose crystal structure is shown in Fig.~\ref{intro}b. Given niobium's heavy atomic mass, Nb ions can be tuned to penetrate to a depth comparable to that of the thin film as opposed to stopping on the surface. Our 90 nm (112)-oriented Cd$_3$As$_2$ thin films were grown via MBE on (111)-oriented GaAs substrates with a buffer layer of 150 nm of (111)-oriented GaSb in between. More details on the growth method of the films can be found in the work by Schumann et al \cite{schumann2016molecular}.

We calculated the electronic band structure of Cd$_3$As$_2$, as shown in Fig.~\ref{intro}c, using density functional theory (DFT) (see details in Supplementary Information I with additional references \cite{kresse1996efficient,kresse1999ultrasoft,perdew1996generalized,wang2021vaspkit}). The Fermi level lies above the Dirac nodes, so a dopant is necessary to shift the level. 

\begin{figure*}[!htbp]
    \centering
    \includegraphics[width=0.85\linewidth]{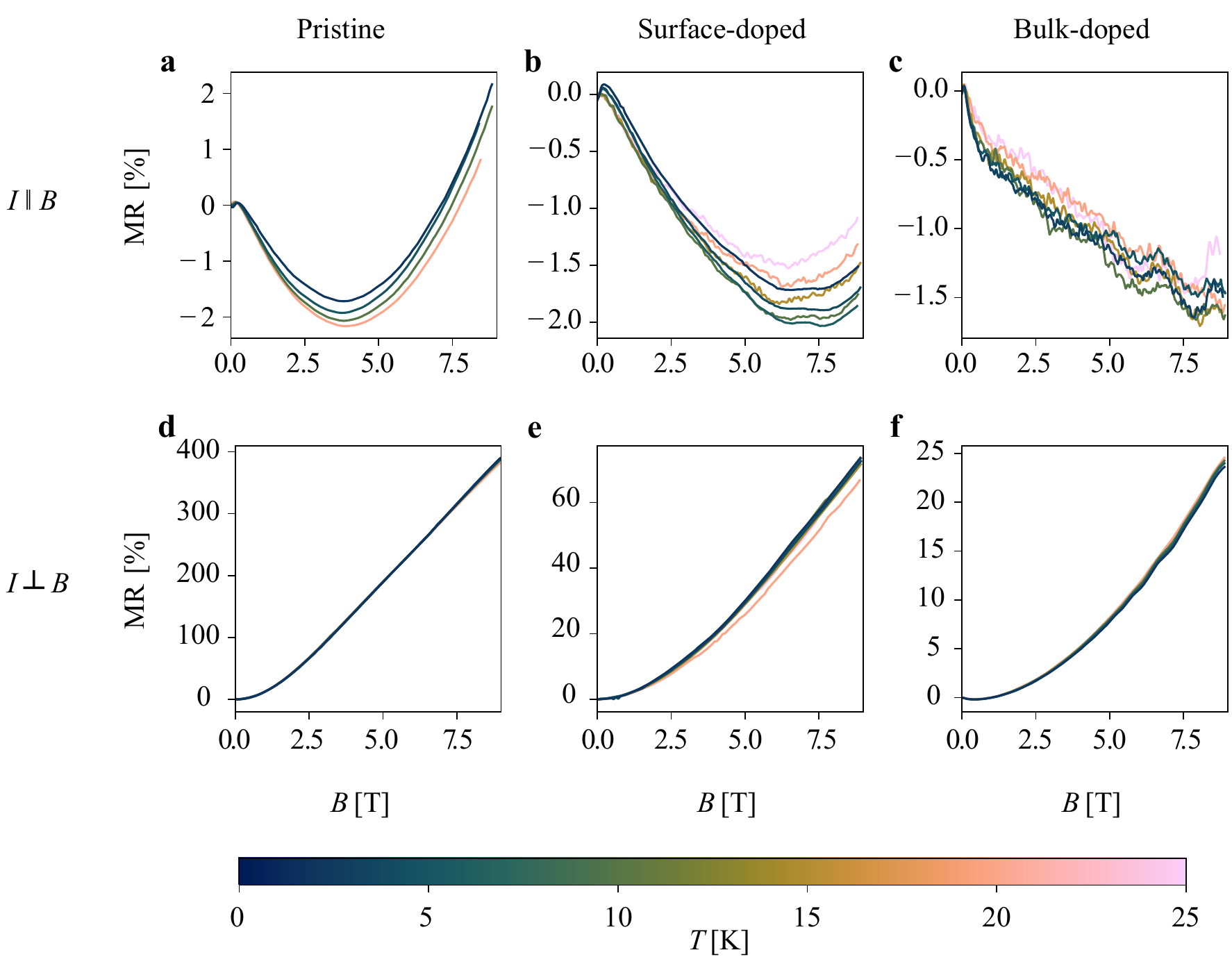}
    \caption{Longitudinal magnetoresistance (MR) for $I \parallel B$ with increasing magnetic field at different temperatures for the \textbf{a.} pristine, \textbf{b.} surface-doped, and \textbf{c.} bulk-doped \ce{Cd3As2} samples. Same for \textbf{d-f.} respectively for $I\perp B$. All samples under $I\parallel B$ conditions exhibit negative longitudinal MR greater than $1\%$. As the magnetic field increases, the longitudinal MR decreases until reaching a minimum and then turns up. The minimum of longitudinal MR occurs near $4$T for the pristine sample, but shifts to around $6-8$T for the surface-doped sample, and over $9$T for the bulk-doped sample, suggesting the chiral charge pumping effects are enhanced with disorder.}
    \label{fig4}
\end{figure*}

In addition to using DFT to estimate the doping concentration of Nb ions, we simulated the irradation depth profiles of Nb bombardment. We used the Stopping Range of Ions in Matter software (SRIM/TRIM) \cite{SRIM} to determine the ion penetration depth within the crystal. We simulated the irradiation of Cd$_3$As$_2$ thin films with 25 keV and 200 keV Nb ions to achieve surface and bulk doping respectively as shown in \figref{fig2}a. The detailed calculations are described in Supplementary Information II. To experimentally confirm the irradation depth profiles, we performed secondary ion mass spectrometry (SIMS) analysis \cite{SIMS} of the irradiated samples in collaboration with the SIMS Lab of Beamline Inc. \cite{SIMSLab}. We measured the Nb concentrations as a function of depth using an O beam as shown in \figref{fig2}b and show agreement with the SRIM/TRIM calculations. The 25 keV irradiation dopes the majority of Nb atoms within 400Å of the surface of Cd$_3$As$_2$ while the 200 keV irradiation dopes the bulk of the Cd$_3$As$_2$ thin film. 

We confirmed the structural integrity of the samples following ion implantation scanning transmission electron microscopy (STEM). As shown in Fig.~\ref{fig3}, STEM imaging verifies the overall crystallinity of the \ce{Cd3As2}/GaSb/GaAs heterostructure. Fig.~\ref{fig3}a, b are low and middle resolution STEM images viewed along [$1\bar{1}0$] direction of the Nb-Cd$_3$As$_2$ acquired by high-angle annular dark-field (HAADF). Fig.~\ref{fig3}a shows that most of the structural damage is concentrated in the GaSb buffer layer whereas the Cd$_3$As$_2$ layer remains structurally intact. High-resolution images along the [$1\bar{1}0$] zone axis (see Fig.~\ref{fig3}c) show well-defined atomic periodicity, consistent with the simulated STEM-HAADF image (see Fig.~\ref{fig3}e) calculated based on the ideal Cd$_3$As$_2$ crystal structure \cite{zhu2023interface}. A careful comparison, however, reveals relative intensity differences between the experimental and simulated images in some atomic columns, as indicated by Fig.~\ref{fig3}d-f. These discrepancies in relative intensities indicate isolated point defects likely due to ion bombardment. Our STEM observations confirm that ion implantation introduces minimal structural disruption while enhancing NLMR.

To probe the effects of ion implantation on NLMR, we performed systematic magnetoresistance (MR) measurements using a Quantum Design Physical Property Measurement System (PPMS) across the pristine, surface-doped (25 keV), and bulk-doped (200 keV) samples under varying magnetic field strengths (0–9 T) and orientations (longitudinal ($I\parallel B$) and transverse ($I\perp B$)) as illustrated in \figref{fig4}. The details of the data collection of these measurements are discussed in Supplementary Information III with an additional reference \cite{son2013chiral}.
In the longitudinal setup, all samples exhibit a pronounced NLMR, a known transport signature of the ABJ chiral anomaly \cite{son2013chiral,xiong2015evidence,zhang2016signatures,huang2015observation}, exceeding $-1$\% as shown in \figref{fig4}a--c. For the pristine sample, we observe that magnetoresistance reaches its minimum near 4 T as shown in \figref{fig4}a. Interestingly, Nb doping does not suppress this feature as often expected with disorder. In the surface-doped sample, the magnetoresistance minimum shifts to 7–8 T (see \figref{fig4}b) while in the bulk-doped sample the minimum is delayed to above 9 T (see \figref{fig4}c). This progressive shift indicates that deeper or more homogeneous disorder enhances the NLMR, delaying its suppression to higher magnetic fields. A theoretical model supporting this observation is provided in the Supplementary Information IV.

We have demonstrated a practical approach to enhance the NLMR in the Dirac semimetal Cd$_3$As$_2$ via high-energy Nb implantation enabled by accelerator-based techniques. To our knowledge, fast ion implantation has rarely been applied to enhance quantum properties. Transport measurements reveal an enhancement of NLMR with surface and bulk Nb implantation, potentially indicating enhanced chiral transport behavior through disorder. This implantation method offers a highly controllable and broadly applicable strategy for understanding and tuning chiral properties in quantum materials.

\section*{ACKNOWLEDGEMENTS}
The authors MM, ER, AC, and DCC contributed equally to this work. MM and AC acknowledge support from the US Department of Energy (DOE), Office of Science (SC), Basic Energy Sciences (BES), Award No. DE-SC0020148. \textcolor{myred}{Research at BNL was supported by US DOE/BES, Materials Sciences and Engineering Division under Contract No. DE-SC0012704.} ER is supported by NSF Designing Materials to Revolutionize and Engineer our Future (DMREF) Program with Award No. DMR-2118448. DCC is supported by the NSF Convergence Accelerator Award ITE-2345084. ML acknowledges the support from the Class of 1947 Career Development Professor Chair, and the support from R. Wachnik.  

\bibliographystyle{apsrev4-2}
\bibliography{references,refs_new}
\end{document}


\title{Tuning chiral anomaly signature in a Dirac semimetal via fast-ion implantation: Supplementary Information}

\author{Manasi Mandal}
\email[]{manasim@mit.edu}
\affiliation{Quantum Measurement Group, MIT, Cambridge, MA 02139, USA}
\affiliation{Department of Nuclear Science and Engineering, MIT, Cambridge, MA 02139, USA}
\thanks{These authors contribute to this work equally.}

\author{Eunbi Rha}
\thanks{These authors contribute to this work equally.}
\affiliation{Quantum Measurement Group, MIT, Cambridge, MA 02139, USA}
\affiliation{Department of Nuclear Science and Engineering, MIT, Cambridge, MA 02139, USA}

\author{Abhijatmedhi Chotrattanapituk}
\thanks{These authors contribute to this work equally.}
\affiliation{Quantum Measurement Group, MIT, Cambridge, MA 02139, USA}
\affiliation{Department of Electrical Engineering and Computer Science, MIT, Cambridge, MA 02139, USA}

\author{Denisse C\'{o}rdova Carrizales}
\thanks{These authors contribute to this work equally.}
\affiliation{Quantum Measurement Group, MIT, Cambridge, MA 02139, USA}
\affiliation{Department of Nuclear Science and Engineering, MIT, Cambridge, MA 02139, USA}

\author{Alexander Lygo}
\affiliation{Materials Department, University of California, Santa Barbara, CA 93106, USA}

\author{Kevin B. Woller}
\affiliation{Department of Nuclear Science and Engineering, MIT, Cambridge, MA 02139, USA}
\affiliation{Plasma Science and Fusion Center, Massachusetts Institute of Technology, Cambridge, MA 02139, USA}

\author{Mouyang Cheng}
\affiliation{Quantum Measurement Group, MIT, Cambridge, MA 02139, USA}
\affiliation{Center for Computational Science and Engineering, MIT, Cambridge, MA 02139, USA}
\affiliation{Department of Materials Science and Engineering, MIT, Cambridge, MA 02139, USA}

\author{Ryotaro Okabe}
\affiliation{Quantum Measurement Group, MIT, Cambridge, MA 02139, USA}
\affiliation{Department of Chemistry, MIT, Cambridge, MA 02139, USA}

\author{Guomin Zhu}
\affiliation{Materials Department, University of California, Santa Barbara, CA 93106, USA}

\author{Kiran Mak}
\affiliation{Quantum Measurement Group, MIT, Cambridge, MA 02139, USA}
\affiliation{Department of Materials Science and Engineering, MIT, Cambridge, MA 02139, USA}
\affiliation{Department of Physics, MIT, Cambridge, MA 02139, USA}

\author{Chu-Liang Fu}
\affiliation{Quantum Measurement Group, MIT, Cambridge, MA 02139, USA}
\affiliation{Department of Nuclear Science and Engineering, MIT, Cambridge, MA 02139, USA}

\author{Chuhang Liu}
\affiliation{Condensed Matter Physics and Materials Science Department, Brookhaven National Laboratory, Upton, NY 11973, USA}

\author{Lijun Wu}
\affiliation{Condensed Matter Physics and Materials Science Department, Brookhaven National Laboratory, Upton, NY 11973, USA}

\author{Yimei Zhu}
\email[]{zhu@bnl.gov}
\affiliation{Condensed Matter Physics and Materials Science Department, Brookhaven National Laboratory, Upton, NY 11973, USA}

\author{Susanne Stemmer}
\email[]{stemmer@mrl.ucsb.edu}
\affiliation{Materials Department, University of California, Santa Barbara, CA 93106, USA}

\renewcommand{\thefootnote}{\fnsymbol{1}}
    
\author{Mingda Li}
\email[]{mingda@mit.edu}
\affiliation{Quantum Measurement Group, MIT, Cambridge, MA 02139, USA}
\affiliation{Department of Nuclear Science and Engineering, MIT, Cambridge, MA 02139, USA}
\affiliation{Center for Computational Science and Engineering, MIT, Cambridge, MA 02139, USA}

\maketitle
\tableofcontents
\clearpage

\section{Density Functional Theory (DFT) Calculations}
The electronic properties of Cd$_3$As$_2$ are calculated using the Vienna ab initio simulation package (VASP)\,\cite{kresse1996efficient}. The Projector-Augmented-Wave method\,\cite{kresse1999ultrasoft} and Perdew-Burke-Ernzerhof\,\cite{perdew1996generalized} exchange-correlation functional are used for all DFT calculations. The calculations are performed on a $3\times3\times3$ K-point mesh centered at Gamma point with encut energy value as 400 eV. The SCF single-point electronic energy convergence criterion is set to 1$\times$10$^{-6}$ eV, and the structural relaxation is performed until the residual forces are below 0.01 eV/$\overset{\circ}{\mathrm{A}}$.
The electronic band structure calculation is assisted by the VASPKIT package\,\cite{wang2021vaspkit}.

\section{SRIM and TRIM calculations}
To simulate the experimental ion irradiation and determine the dopant concentration profiles, we employed the SRIM (Stopping and Range of Ions in Matter) and TRIM (Transport of Ions in Matter) software \cite{SRIM}. For this experiment, we simulated $25$ keV and $200$ keV Nb ion bombardment of a $90$ nm Cd$_3$As$_2$ layer on a $150$ nm GaSb substrate. From this we obtained a 3d dopant concentration profile from both the incident ions and subsequent scattering events. During the experiment, we irradiated the $5\times5$ mm$^2$ sample with an 5 mm diameter ion beam with fluctuating beam current in orders of tens of nA for 6 hours with the energy of 25 keV for surface doping, and for 24 hours with the energy of 200keV for bulk doping, respectively. The ion energies were chosen to yield a sample with either surface or bulk doping with the same target doping concentration $\approx10^{15}$ cm$^{-2}$. We approximate the number of implanted ions from the incident ion flux, dependent on the beam current and aperture size, and irradiation time, accounting for the separation of secondary electrons and subsequent damage cascades.

\section{Magnetic Field Dependence of Resistivity}
\begin{figure*}[ht!]
    \centering
    \begin{tikzpicture}
        \node[anchor=north west] (image) at (0,0) 
        {\includegraphics[width=7in]{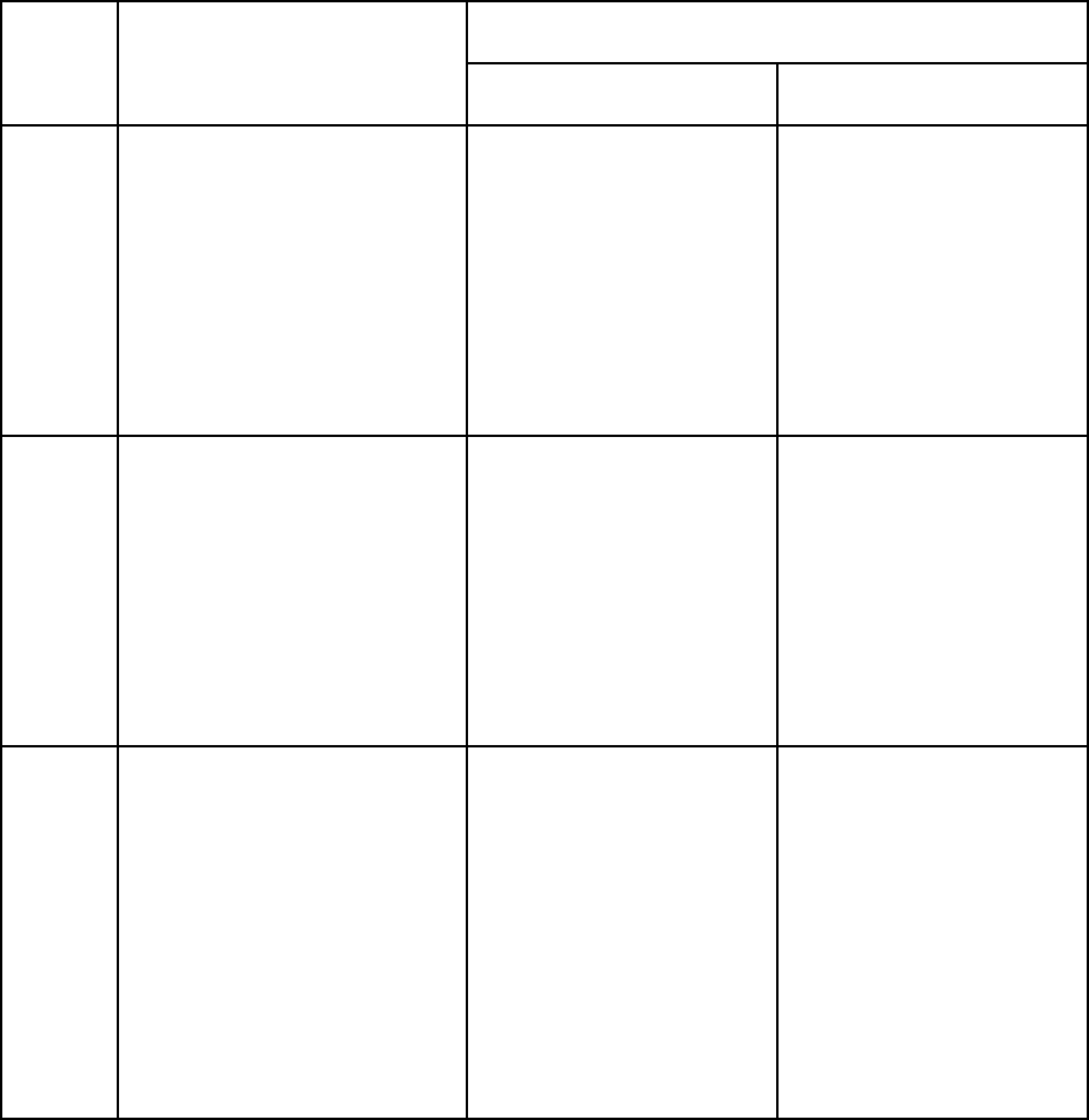}};
        \node[anchor=north west] (image) at (0.95in,-1in) 
        {\includegraphics[height = 1.7in]{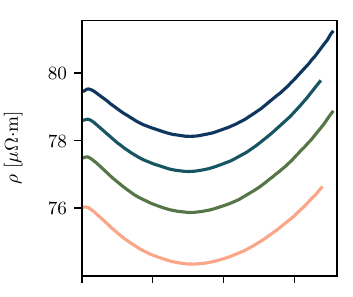}};
        \node[anchor=north west] (image) at (3.1in,-1in) 
        {\includegraphics[height = 1.7in]{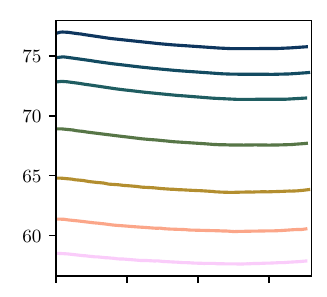}};
        \node[anchor=north west] (image) at (5.1in,-1in) 
        {\includegraphics[height = 1.7in]{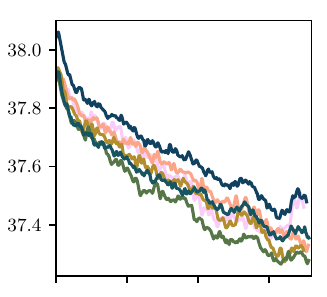}};
        \node[anchor=north west] (image) at (0.95in,-3in) 
        {\includegraphics[height = 1.7in]{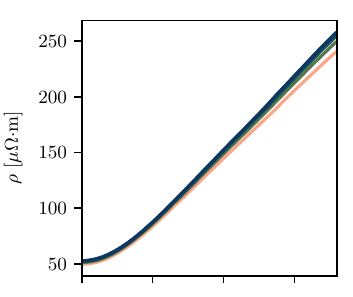}};
        \node[anchor=north west] (image) at (3.1in,-3in) 
        {\includegraphics[height = 1.7in]{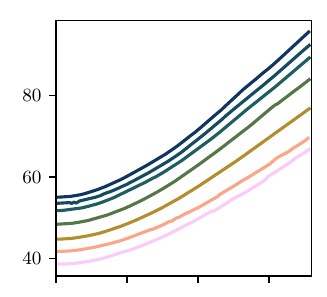}};
        \node[anchor=north west] (image) at (5.1in,-3in) 
        {\includegraphics[height = 1.7in]{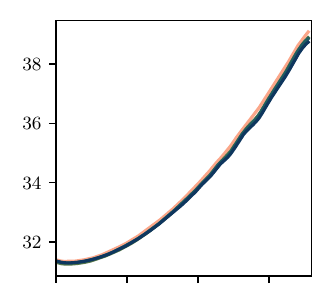}};
        \node[anchor=north west] (image) at (0.95in,-5in) 
        {\includegraphics[height = 2in]{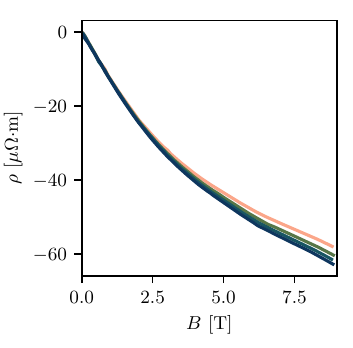}};
        \node[anchor=north west] (image) at (3.1in,-5in) 
        {\includegraphics[height = 2in]{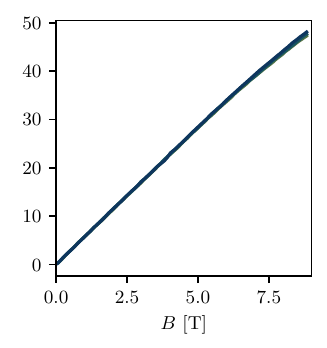}};
        \node[anchor=north west] (image) at (5.1in,-5in) 
        {\includegraphics[height = 2in]{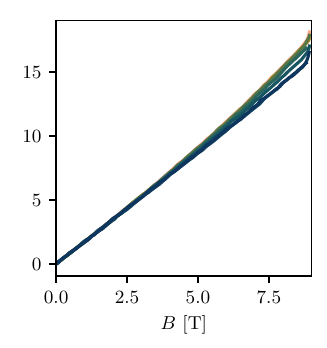}};

        \node[anchor=north west] (annotate) at (1.7in,-0.4in) {Pristine};
        \node[anchor=north west] (annotate) at (4.8in,-0.15in) {Doped};
        \node[anchor=north west] (annotate) at (3.8in,-0.57in) {Surface};
        \node[anchor=north west] (annotate) at (5.85in,-0.57in) {Bulk};
        \node[anchor=north west] (annotate) at (0.05in,-0.35in) {Resistivity};
        \node[anchor=north west] (annotate) at (0.15in,-1.6in) {\begin{tabular}{c}
        $\rho_{xx}$ \\
        $I\parallel B$
        \end{tabular}};
        \node[anchor=north west] (annotate) at (0.15in,-3.5in) {\begin{tabular}{c}
        $\rho_{xx}$ \\
        $I\perp B$
        \end{tabular}};
        \node[anchor=north west] (annotate) at (0.15in,-5.8in) {\begin{tabular}{c}
        $\rho_{xy}$ \\
        $I\perp B$
        \end{tabular}};

        \node[anchor=north west] (a) at (1.1in,-0.9in) {\textbf{a}};
        \node[anchor=north west] (b) at (3.1in,-0.9in) {\textbf{b}};
        \node[anchor=north west] (c) at (5.1in,-0.9in) {\textbf{c}};
        \node[anchor=north west] (d) at (1.1in,-2.9in) {\textbf{d}};
        \node[anchor=north west] (e) at (3.1in,-2.9in) {\textbf{e}};
        \node[anchor=north west] (f) at (5.1in,-2.9in) {\textbf{f}};
        \node[anchor=north west] (g) at (1.1in,-4.9in) {\textbf{g}};
        \node[anchor=north west] (h) at (3.1in,-4.9in) {\textbf{h}};
        \node[anchor=north west] (i) at (5.1in,-4.9in) {\textbf{i}};
    \end{tikzpicture}
    \begin{tikzpicture}
        \node[anchor=north west] (image) at (0,0) 
        {\includegraphics{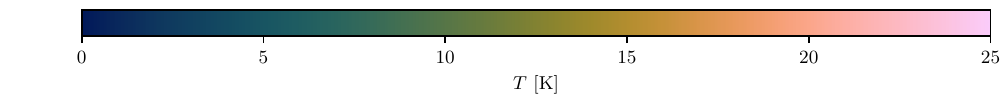}};
    \end{tikzpicture}
    \vspace*{-15mm}
    \caption{Variation of the longitudinal resistivity with the magnetic field parallel to current for the \textbf{a.} pristine, \textbf{b.} surface-doped, and \textbf{c.} bulk-doped sample. Same for \textbf{d-f.} with magnetic field perpendicular to the current direction and \textbf{g-i.} for transverse resistivity with magnetic field perpendicular to the current direction. The opposite slope of the pristine case \textbf{g.} is due to the shift in majority charge carriers from electrons to holes with doping.}
    \label{exp_analysis_summary}
\end{figure*}

The resistivity measurements of the samples were carried out using the electrical transport option (ETO) of the physical property measurement system (PPMS). To minimize the influence of contact misalignment, we utilized the symmetric probe procedure in which the longitudinal resistivity, $\rho_{xx}$, exhibited symmetry while the transverse resistivity, $\rho_{xy}$, displayed anti-symmetry when the magnetic field is in the plane containing both current, and sample normal. The symmetric probe can be done by utilizing the following symmetrize equations:
\begin{equation}
    \rho_{xx}(B)=\dfrac{\rho_{xx}(+B)+\rho_{xx}(-B)}{2}, \rho_{xy}(B)=\dfrac{\rho_{xy}(+B)-\rho_{xy}(-B)}{2}.
\end{equation}
The magnetic field dependence resistivity of each experimental setup at different temperatures are illustrated in Fig.~\ref{exp_analysis_summary}. The magnetoresistance in the main text is calculated according to the following equation,
\begin{equation}
    \text{MR}[\%] = \dfrac{\rho_{xx}(B)-\rho_{xx}(0)}{\rho_{xx}(0)}\times 100
\end{equation}

\section{A Simple Theoretical Model of Enhanced Chiral Anomaly}

To gain more theoretical insights on the defect-tunable Chiral anomaly, we note that for $I \parallel B$, the longitudinal conductivity is contributed by both the Chiral anomaly term $\sigma^A_{xx}$ and the normal Drude term $\sigma^D_{xx}$, namely $\sigma_{xx}(B) = \sigma^A_{xx}(B) + \sigma^D_{xx}(B)$.
As is derived by Son and Spivak \cite{son2013chiral}, the Chiral anomaly-related contribution scales quadratically with magnetic field $B$,
\begin{equation}
    \sigma^A_{xx}(B) = \frac{e^2}{4 \pi^2 \hbar c} \frac{v}{c} \frac{(e B)^2 v^2}{\mu^2} \tau,
\end{equation}
here $v$ is the quasi-particle velocity near the Dirac cone, $\tau$ is the relaxation time for electric intervalley scattering, and $\mu$ is the Fermi level with respect to the Dirac cone.
As for the Drude term, at small $B$ it decreases quadratically with $B$ field,
\begin{equation}
    \sigma_{xx}^D(0)-\sigma_{xx}^D(B) \propto \sigma_{xx}^D(0)\left(\omega_c \tau_f\right)^2 .
\end{equation}
where $\omega_c =eB/m$, and $\tau_f$ is the mean free time for Drude ionic scattering.
In metallic \ce{Cd3As2} we have $\tau \gg \tau_f$, so the chiral anomaly contribution dominates, making the conductivity increase quadratically. Thus the longitudinal resistivity scales as
\begin{equation}
    \rho_{xx}(B) \sim \sigma_{xx}^{-1}(B) = \frac{1}{\sigma_{xx}^D(0) + \lambda B^2}
\end{equation}
with $\lambda>0$. As the magnetic field further increases, the Drude contribution deviates from the quadratic scaling first, introducing a 4$^\text{th}$ power contribution,
\begin{equation}
    \rho_{xx}(B) \sim \sigma_{xx}^{-1}(B) = \frac{1}{\sigma_{xx}^D(0) + \lambda B^2 - \xi B^4}
\end{equation}
with $\lambda,\xi>0$. Under these approximations, the turning point of magnetic field for minimal MR is $B_c=\sqrt{\lambda/2\xi}$. Now as we dope the bulk \ce{Cd3As2}, dopants tune the Fermi level closer to the Dirac cone, i.e. $\mu$ gets smaller. This increases the proportionality constant $\lambda$, and thus increases the critical magnetic field $B_c$ for the chiral anomaly behavior.

\newpage
\bibliography{references.bib,refs_new.bib}